\documentclass[%
 aip,
rsi,%
 amsmath,amssymb,
 reprint,%
]{revtex4-1}

\usepackage{graphicx}
\usepackage{dcolumn}
\usepackage{bm}

\begin{document}

\preprint{AIP/123-QED}

\title{Terahertz conductivity of the magnetic Weyl semimetal Mn$_{3}$Sn films}

\author{Bing Cheng}

\author{Youcheng Wang}

\author{D. Barbalas}

\affiliation{The Institute of Quantum Matter, Department of Physics and Astronomy, The Johns Hopkins University, Baltimore, Maryland 21218, USA}

\author{Tomoya Higo}
\affiliation{Institute for Solid State Physics, University of Tokyo, Kashiwa, 227-8581, Japan}

\author{S. Nakatsuji}
\affiliation{The Institute of Quantum Matter, Department of Physics and Astronomy, The Johns Hopkins University, Baltimore, Maryland 21218, USA}
\affiliation{Institute for Solid State Physics, University of Tokyo, Kashiwa, 227-8581, Japan}

\author{N. P. Armitage}
\affiliation{The Institute of Quantum Matter, Department of Physics and Astronomy, The Johns Hopkins University, Baltimore, Maryland 21218, USA}
\affiliation{Japan Society for the Promotion of Science, International Research Fellow, Institute for Solid State Physics, The University of Tokyo, Kashiwa 277-8581, Japan}

\date{\today}

\begin{abstract}
Mn$_{3}$Sn is a non-collinear antiferromagnet which displays a large anomalous Hall effect at room temperature.  It is believed that the principal contribution to its anomalous Hall conductivity comes from Berry curvature. Moreover, dc transport and photoemission experiments have confirmed that Mn$_{3}$Sn may be an example of a time-reversal symmetry breaking Weyl semimetal.   Due to a small, but finite moment in the room temperature inverse triangular spin structure, which allows control of the Hall current with external field, this material has garnered much interest for next generation memory devices and THz spintronics applications.  In this work, we report a THz range study of oriented Mn$_{3}$Sn thin films as a function of temperature.  At low frequencies we found the optical conductivity can be well described by a single Drude oscillator. The plasma frequency is strongly suppressed in a temperature dependent fashion as one enters the 260 K helical phase.   This may be associated with partial gapping of the Fermi surfaces that comes from breaking translational symmetry along the c-axis.  The scattering rate shows quadratic temperature dependence below 200 K, highlighting the possible important role of interactions in this compound.

\end{abstract}
               
\maketitle

The anomalous Hall effect (AHE) conventionally occurs in ferromagnetic metals (FMs) and is usually interpreted as an effect beyond the ordinary Hall effect that arises from spontaneous magnetization\cite{AHE_2010}.   It can arise based on three different mechanisms: side jump scattering, skew scattering, and intrinsic effects.   The latter arises through Berry curvature.  This Berry curvature is determined by the geometry of Bloch wavefunctions and acts  in many ways as an effective magnetic field in momentum space.  Recently, the possibility of an intrinsic AHE in systems without net magnetization has been explored, such as in spin liquid and antiferromagnetic systems\cite{Machida10,AHE_Mn3Sn,AHE_MnGe_2016,Mn3Ge_AHE}. One of the most promising possibilities are non-collinear antiferromagnets.  In most antiferromagnets with collinear spins, the Berry curvature is zero. However theory has found that in a triangular spin structures with non-collinear moments a Berry curvature can be present\cite{AHE_noncollinear_2014,AHE_topology_2017}.  One of the most likely realizations for this physics is Mn$_{3}$Sn\cite{AHE_Mn3Sn}, which has attracted recent attention because of its unique magnetic structure and symmetries. Below T$_{N}$ $\sim$ 420 K, it undergoes an antiferromagnetic transition and enters a non-collinear inverse triangular spin state\cite{Mn3Sn_spin_structure_1986,Mn3Sn_spin_structure_1987} that shows a large anomalous Hall conductivity, Nernst effect, and magneto-optical Kerr effect (MOKE) \cite{AHE_Mn3Sn,Mn3Sn_Nernst_2017,Mn3Sn_Kerr_2018}.  The system transitions to a helical (spiral) spin structure on cooling below 260 K in which the AHE is quenched \cite{sung2018magnetic,Mn3Sn_spin_structure_1987,Mn3Sn_Helical_neutron_1993}.

First-principle band structure calculations of Mn$_{3}$Sn showed multiple pairs of Weyl nodes and three-dimensional linear dispersions in momentum space that have been confirmed by recent photoemission experiments\cite{Mn3Sn_photoemission_2017}.   Mn$_{3}$Sn is of particular applications interest because in the non-collinear inverse triangular spin state, the system also exhibits a very small net magnetic moments of ~3 m$\mu_B$/Mn that allows the magnetic state and the AHE to be controlled and manipulated with external field, while giving negligible fringe field itself.  This makes this system of interest for next generation memory devices and THz spintronics.    In this regard THz dynamics of Mn$_{3}$Sn are particularly important for applications.

In this work, we take advantage of recent technical advances in thin film preparation \cite{higo2018anomalous} to present the first THz range optical study of a polycrystalline Mn$_{3}$Sn thin film. We found the complex optical conductivity can be well reproduced by a single Drude oscillator, indicative of excellent metallic conduction. The plasma frequency of the Drude oscillator is suppressed with lowering temperature below 200 K, which strongly hints at the partial gap opening below the phase transition to the helical phase at 260 K.  The scattering rate has an approximately quadratic dependence on temperature at low temperature, before crossing over to a weaker temperature dependence above the 250 K scale.  These features enable us to set the upper limits of the effective mass and mobility for free carriers and supports the notion of a correlated Weyl phase in Mn$_{3}$Sn.

\begin{figure*}[t]
\includegraphics[clip,width=5.5in]{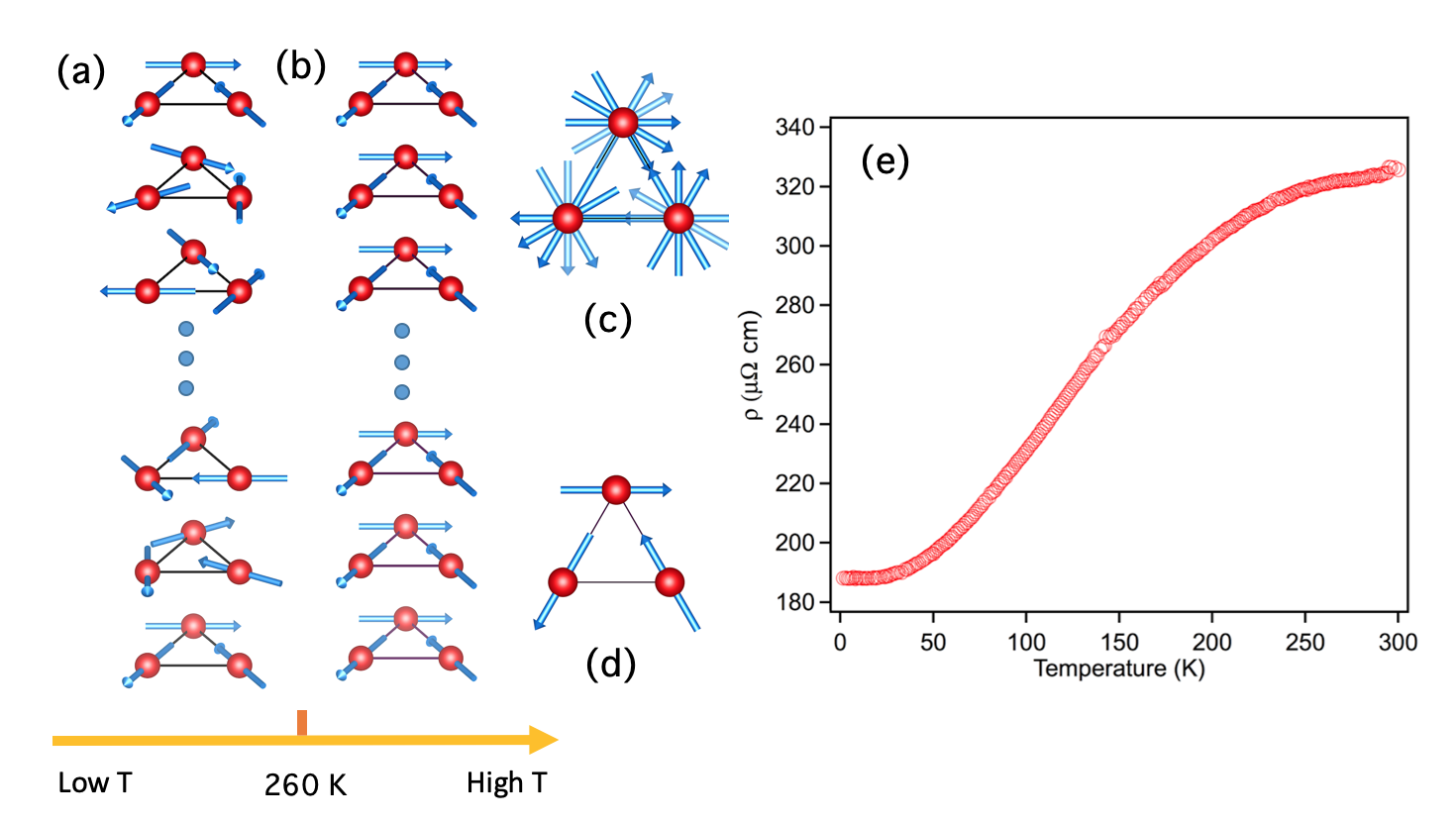}

\caption{(Color online) Magnetic structure of Mn$_{3}$Sn film in (a) magnetic helical phase below 260 K and (b) magnetic inverse triangular phase above 260 K. Top view of (c) helical phase and (d) inverse triangular phase. (e) dc resistivity of polycrystalline Mn$_{3}$Sn film. }
\label{xxx}

\end{figure*}

Polycrystalline Mn$_{3}$Sn films were deposited to a thickness of 100 nm by DC magnetron sputtering at room temperature onto Si/SiO$_2$ substrates from a Mn$_{2.5}$Sn target. The sputtering power and Ar gas pressure are 60 W and 0.3 Pa.  The film is post-deposition annealed at 500$^\circ$C for 1 h. The annealing procedure after the deposition crystallizes the as-deposited amorphous film into a polycrystalline form of Mn$_3$Sn.  Further details of the deposition and sample treatment can be found in \cite{higo2018anomalous}.  XRD revealed that the almost randomly oriented polycrystalline Mn$_3$Sn film.   This has been shown recently to allow the large anomalous Hall effect comparable to the bulk Mn$_3$Sn. \cite{higo2018anomalous}.

Complex values of the THz transmission of these films were measured in a home-built THz time-domain spectroscopy (TDTS) spectrometer with a closed-cycle 7 T superconducting magnet.  The complex conductivity can be extracted by using the appropriate expression in the thin film approximation: $T(\omega) = \frac{1+n}{1+n+Z_{0}d\sigma(\omega)}exp[\frac{i\omega}{c}(n-1)\Delta L]$\cite{Bing16_1}. Here $T(\omega)$ is the complex transmission function as referenced to a bare  substrate, $\sigma$($\omega$) is the complex optical conductivity, $d$ is the thickness of the film, and $n$ is the index of refraction of the substrate. $\Delta L$ is the small thickness difference between samples and reference substrates, and $Z_{0}$ is the vacuum impedance, which is approximately 377 $\Omega$.

Fig. 1(a) to (d) depict the magnetic structure schematically of Mn$_{3}$Sn.  As mentioned above, below a Neel temperature T$_N$ = 420 K, Mn$_{3}$Sn orders in an inverse triangular spin configuration with negative vector chirality in the kagome lattice [Fig. 1(b)]. Mn moments lie in the ab plane and form 120$^\circ$ angles with each other [Fig. 1(d)]. In each magnetic primitive unit cell, there are two triangles along the vertical direction related by inversion symmetry. In this phase, a very large anomalous Hall conductivity is observed at zero field which originates from non-zero Berry curvature in momentum space induced by the cluster multipole order\cite{Cluster_Multipole_17}. Furthermore, Weyl nodes in the bulk and Fermi arcs on the surface was predicted by band structure calculations in this phase, which have been confirmed by photoemission measurements\cite{Mn3Sn_photoemission_2017}.  dc transport displays a negative magnetoresistance with $B$ $\parallel$ $I$, which is regarded as a signature of the chiral anomaly that describes the breakdown of chiral symmetry in Weyl semimetals when presenting parallel electric and magnetic fields\cite{Mn3Sn_photoemission_2017}. Our previous study has found that, below 260 K, the Mn$_{3}$Sn thin film undergoes a magnetic phase transition to the helical magnetic state\cite{higo2018anomalous} [Fig. 1(a)]. For each triangle, the Mn moments still lie in $ab$ plane with 120$^\circ$ pattern, but the moment of each triangle is rotated by an angle about the vertical direction forming a helical structure [Fig. 1(c)]. In this phase, the anomalous Hall conductivity at zero field vanishes. Both of the magnetic states are metallic. Fig. 1(e) shows that the dc resistivity $\rho$ increases with increasing temperature. Above 250 K, the resistivity seems to gradually saturate.   This can be either because the rate of increase of the carrier density decreases or the rate of increase of the scattering rate decreases.  Optical conductivity is a powerful method to sort out these possibilities.   We show below that both play a role.

\begin{figure*}[t]
	\includegraphics[clip,width=7in]{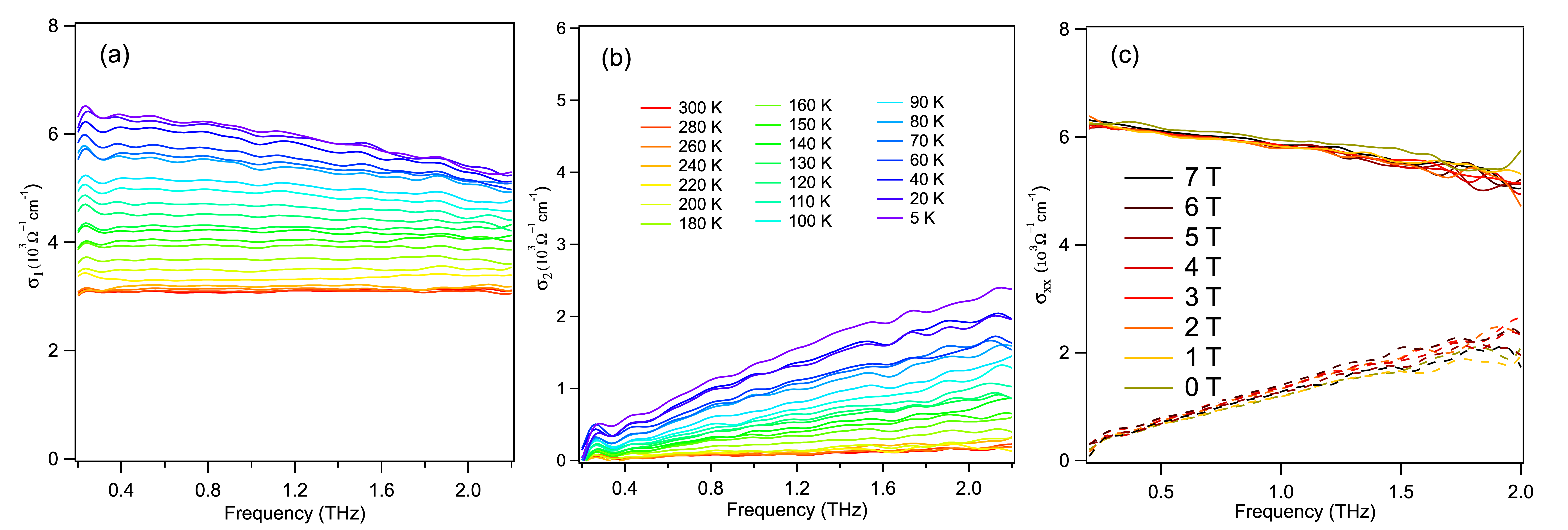}
	\caption{(Color online) (a) Real and (b) imaginary parts of optical conductivity of Mn$_{3}$Sn film at different temperatures. (c) Real and imaginary parts of magneto-terahertz conductivity in Faraday geometry at 6 K. }
	\label{xxx}
\end{figure*}

Figs. 2(a) and (b) show the real and imaginary parts of the optical conductivity measured at different temperatures.  At the low temperature of 5 K, the real part of the optical conductivity $\sigma_{1}$ displays a well-defined conductivity peak centered at zero frequency with $\sigma_{2}$ is an increasing function of frequency, indicative of good metallic behavior. With increasing temperature, the peak becomes broader and $\sigma_{2}$ flattens.   When further increasing temperature above 250 K, $\sigma_{1}$ becomes flat and show weak temperature dependence. In the THz region, no phonons or other absorptions are observed. In this regard, we can use a simple Drude model to fit the optical conductivity:

\begin{equation}
\sigma(\omega)=\epsilon_0[-\sum_{k=1}^{s}{{\omega_{pk}^2}\over{i\omega-\Gamma_{pk}}}-i(\epsilon_\infty-1)\omega].
\label{chik}
\end{equation}

\begin{figure*}[t]
	\includegraphics[clip,width=7in]{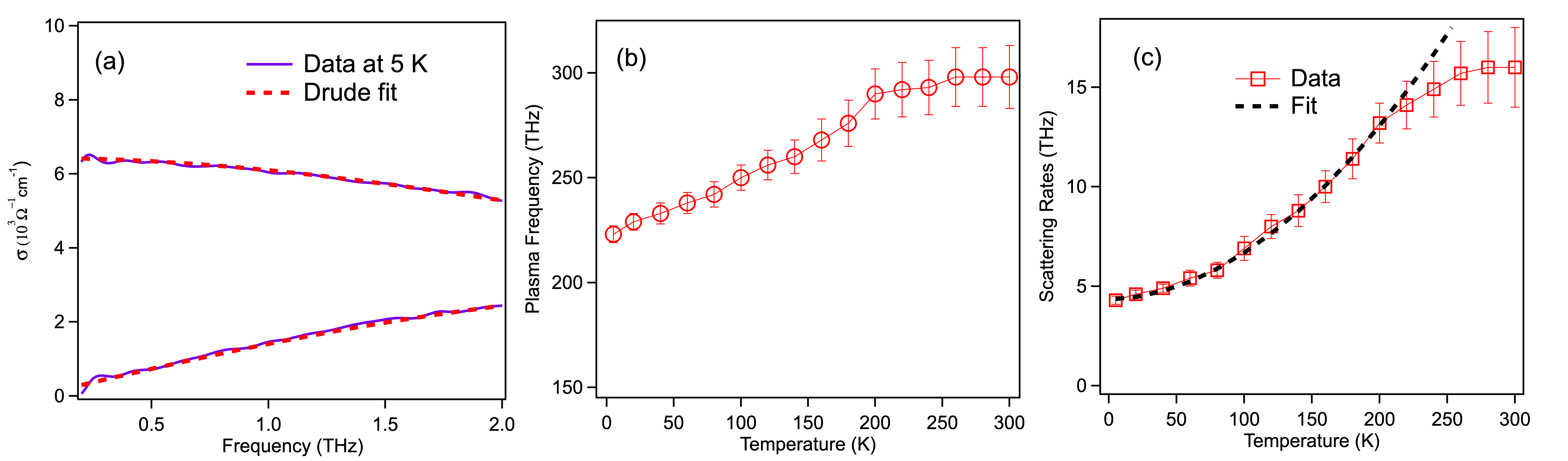}
	\caption{(Color online) (a) Drude fit for real and imaginary parts of terahertz conductivity at 6 K. (b) Temperature dependent plasma frequency. (c) Temperature dependent scattering rates and its power-law fit. }
	\label{xxx}
\end{figure*}

 \noindent Here, $\epsilon_\infty$ represents a background polarizability that originates from absorptions above the measured spectral range including phonons and electronic interband transitions. $s$ is an index that represents a possible sum over a number of the Drude oscillators.  An example fit to the data at 5K is shown in Fig. 3(a). One can see that both the real and imaginary parts of the optical conductivity at 5 K can be well simulated by a single Drude oscillator.
 
 The temperature dependent plasma frequency $\omega_{p}$/2$\pi$ is displayed in Fig. 3(b). One can see that $\omega_{p}$/2$\pi$ has very strong temperature dependence. From 5 K to $\sim$ 200 K, the plasma frequency increases quickly with temperature, but above 200 K it saturates. The plasma frequency is determined by carrier density and effective mass via the expression $\omega_p^2=\frac{ne^{2}}{ \epsilon_0 m^{*}}$. A decreasing plasma frequency indicates the carrier density decreases or effective mass of carriers increases, or both of these effects play partial roles. As shown in Fig. 1(a) and (c), below 250 K Mn$_{3}$Sn film enters a helical magnetic state.  In this state, along the $c$ axis, the magnetic moment \textbf{m$_a$} (along a axis) or \textbf{m$_b$} (along b axis) of Mn varies approximately as a sinusoidal function of the $c$ axis.   This may cause partial gapping of the Fermi surface due to Fermi surface nesting\cite{SDW_review_1988}. In this regard, the decreasing plasma frequency with cooling may hint to the possible partial gap opening as temperature is lowered.  A recent LDA calculation shows that in the helical phase, not only are Weyl points annihilated, but gaps open in the band structure in some regions of momentum space, and in other regions  very flat bands form that should have small spectral weight\cite{Mn3Sn_LDA_2018}.  This is consistent with the decreasing plasma frequency with lowering temperature.

Fig 3(c) shows the scattering rate ($\Gamma/ 2 \pi $) as a function of temperature. The overall trend is that the scattering rate decreases as temperature is lowered. However, the behavior above and below $\sim$ 250 K is quite different. Below 200 K, the scattering rates behave in a typical metallic fashion with power law $aT^{n} + b$ describing the data. The temperature exponent $n$ is extracted to be $n$ = 1.9 $\pm$ 0.1. The value of $n$ being close to 2 may indicate a quasi-Fermi-liquid behavior. In contrast, above 200 K, the scattering rate increases more slowly and seems to be saturating above 250 K.  This unusual temperature dependence may be related to the fact that in the inverse triangular magnetic state, the Berry phase of the Weyl fermions may exempt some backscatterings.

We have also performed extensive measurements in perpendicular magnetic field (Faraday geometry) at low temperature, but saw essentially no effect of field in the THz range.   As shown in Fig. 2(c), even at 7 T, $\sigma_{xx}$ still exhibits a textbook Drude form and does not exhibit any signature of a cyclotron resonance peak (CR). At a particular magnetic field, the CR frequency $\omega_c = \frac{eB}{m^*}$  is determined by the cyclotron mass of the charge carriers\cite{LiangWu_phonon_2015}. The absence of CR in the THz region suggests that these carriers have a large cyclotron mass which pushes the CR to  lower frequency. A rough upper limit for CR given by this study is 0.1 THz at 7 T, which means the mass must be larger than 2 electron masses.  This is an extremely large number for semimetals systems close to having Weyl band structure.  The electronic mobility is estimated to be much smaller than 30 cm$^{2}$V$^{-1}$s$^{-1}$.  The large cyclotron mass and low mobility at 6 K strongly indicate even in the high-temperature Weyl phase, the Weyl fermions will have large effective mass and low mobility. Recent photoemission experiment have shown that Mn$_{3}$Sn exhibits a Weyl semimetal phase with notable electronic correlation\cite{Mn3Sn_photoemission_2017}.  Our observation is consistent with this in that correlations will dress the carriers and increase their effective mass.

In conclusion, we have studied the THz-range optical response of Mn$_3$Sn thin films. The system shows a good metallic state in the whole temperature range. Magneto-terahertz conductivities at 6 K do not exhibit field dependence and strongly indicate the quasiparticles have large effective mass and low electronic mobility, consistent with the fact that Mn$_3$Sn is a correlated electron system.   The suppression of the plasma frequency at low temperatures is consistent with the expected gapping of the Fermi surface in the spatially modulated helical phase.

We would like to thank R. Matsunaga for helpful discussions.  The work at JHU was supported as part of the Institute for Quantum Matter, an Energy Frontier Research Center funded by the U.S. DOE, Office of BES under Award Number DE-SC0019331. NPA also acknowledges additional support from the Japan Society for the Promotion of Science, International Research Fellows Program that supported his visit to Tokyo.  Work at ISSP was supported in part by CREST JPMJCR18T3, Japan Science and Technology Agency, by Grants-in-Aid for Scientific Research 16H02209 and Program for Advancing Strategic International Networks to Accelerate the Circulation of Talented Researchers (No. R2604) from the Japanese Society for the Promotion of Science (JSPS), and by Grants-in-Aids for Scientific Research on Innovative Areas (15H05882 and 15H05883) from the Ministry of Education, Culture, Sports, Science, and Technology of Japan.

\bibliography{Quadratic}

\end{document}